# A Coprocessor for Accelerating Visual Information Processing

W. Stechele*), L. Alvado Cárcel**), S. Herrmann*), J. Lidón Simón**)

*) Technische Universität München
**) Universidad Politecnica de Valencia

Walter.Stechele@ei.tum.de

**ABSTRACT**

*Visual information processing will play an increasingly important role in future electronics systems. In many applications, e.g. video surveillance cameras, data throughput of microprocessors is not sufficient and power consumption is too high. Instruction profiling on a typical test algorithm has shown that pixel address calculations are the dominant operations to be optimized. Therefore AddressLib, a structured scheme for pixel addressing was developed, that can be accelerated by AddressEngine, a coprocessor for visual information processing. In this paper, the architectural design of AddressEngine is described, which in the first step supports a subset of the AddressLib. Dataflow and memory organization are optimized during architectural design. AddressEngine was implemented in a FPGA and was tested with MPEG-7 Global Motion Estimation algorithm. Results on processing speed and circuit complexity are given and compared to a pure software implementation. The next step will be the support for the full AddressLib, including segment addressing. An outlook on further investigations on dynamic reconfiguration capabilities is given.*

## 1. Introduction

There are various applications in visual information processing with real-time requirements, e.g. video surveillance and driver assistance. The algorithms for such applications are not standardized, and might never be, but are a topic of ongoing research. Therefore any hardware platform for implementation needs to be flexible enough to support evolving algorithms, but at the same time needs high computational density to meet the requirements.

The focus of this paper is on the design of a coprocessor for visual information processing, where a key technique is video object segmentation [1].

Algorithms for video object segmentation are using many different operations. They can be grouped into high level control operations and low level pixel manipulation. Control requires a high degree of flexibility and thus is good for processor implementation. Pixel manipulation on the other hand requires applying the same operation on many pixels and thus seems to be a good candidate for hardware acceleration. The problem is how to accelerate without loosing flexibility.

As pixel addressing has shown in [2] to be the dominant part in complexity, a structured approach for the definition of addressing schemes was selected as a basis for further optimization. In order to evaluate the potential of hardware acceleration, a coprocessor architecture, called AddressEngine, was developed, which supports a wide range of image analysis algorithms.

Based on instruction level profiling of a video object segmentation algorithm [3] the maximum achievable acceleration with AddressEngine is estimated as a factor of 30, taking into account that all high level parts of the algorithm are executed on the main CPU and only low level operations are executed on AddressEngine. In contrast to a dedicated hardware accelerator, the use of AddressEngine keeps all flexibility and programmability of the algorithm on the main CPU.

The architecture concept of this coprocessor is described in section 2, starting with an overview on the addressing scheme and pixel processing, which are supported by the coprocessor. In section 3 our prototype implementation is described, including dataflow control and memory management. Results on processing speed and complexity on the FPGA are given, as well as a comparison between performances of FPGA versus software implementation of a test algorithm. The outlook presents some ideas for further investigations with this architecture, exploiting dynamic reconfiguration capabilities of FPGAs [4], [5].

## 2. AddressEngine

Our target platform consists of a hardware coprocessor and a software library, where low level pixel operations are executed in a processing unit with high throughput and low power dissipation.

### 2.1 Addressing Scheme

Although there are various different video algorithms, many of them consist of operations, which use only four ways to access the pixel data: Inter addressing, intra addressing, segment addressing, and segment indexed addressing (figure 1). These four types of addressing were implemented in the AddressLib [2].

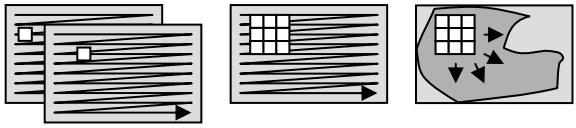

Fig. 1: Pixel addressing schemes: Inter (left), intra (middle), and segment (right) addressing. Arrows indicate the direction of pixel processing.

Inter addressing is the addressing mode where a result for each pixel position is calculated using data from two different frames. Its application may be computation of difference pictures or SAD (Sum of Absolute Differences).

Intra addressing is used in situations where a result is calculated for each pixel as a function of the pixel's original value and the values of its neighbors within the same image. This is typically used for FIR filter like operations, as gradient operators and morphological operators.

Segment addressing is used if arbitrarily shaped segments have to be processed. In this kind of processing, a segment is determined by local neighborhood criteria. First, the pixel processing is done in the same way as for intra addressing. Second, all neighbor pixels which have not been processed before, are tested if they fulfill specified neighborhood criteria. If they do, they might be processed in one of the following steps of the algorithm. By operating this way, an expansion process takes place: Beginning with a set of start pixels, all pixels of the segment are processed in order of geodesic distance.

Segment indexed addressing is an addressing method, which is used in parallel to one of the above addressing methods, when data associated to a segment is needed or generated during the pixel processing, e.g. segment identification numbers. This is done accessing an indexed table.

The first two schemes are well known from frame based or block based video processing. The third addressing scheme is used for pixel addressing of arbitrarily shaped segments in a rectangular frame. The fourth scheme represents indexed table accesses and differs from the other schemes by not addressing pixel data.

## 2.2 Pixel Processing

Pixel-level operations may be separated into basic sub-functions, such as add, sub, mult, grad, in order to achieve efficiency and flexibility. These sub-functions can be combined to form more complex operations, e.g. luminance/chrominance difference between neighboring pixels for homogeneity check, or morphological gradient operations.

## 3. Prototype Implementation

The hardware architecture of the AddressEngine coprocessor includes, in this first version, a subset of the four types of pixel addressing used in the AddressLib software, the intra- and inter addressing modes. The coprocessor has been implemented using the Alpha Data board ADM XRC-II, which contains a VirtexII 3000 FPGA with 216 kBytes of embedded RAM memory, a ZBT SRAM segmented memory (6 Mbytes) made up of 6 independent banks with one write-read 32 bits long port each. The communication between PC and the coprocessor is interrupt oriented and happens through the PCI bus which also has a width of 32 bits. Therefore the architecture design decisions are determined not only by the specifications of the AddressLib software but also by the constraints due to the hardware available in the board used.

The coprocessor architecture deals with both pixel addressing and pixel processing. It is statically configurable in this first version, since the same operation is applied to all the pixels in the whole image for one AddressEngine call. Thus the architecture implements both the processing unit and an input output interface between PC and the processing unit. This is designed to optimize the bus PCI utilization, to obtain an efficient dataflow and to minimize the number of memory accesses by means of pixel reusing and parallel neighbourhood loading.

The general scheme of the architecture, as illustrated in figure 2, consists of the following parts: The on-board memory (ZBT), the intermediate memory system (IIM/OIM), the processing unit, the pixel level controller, transmission units (TxUs), and the image level controller.

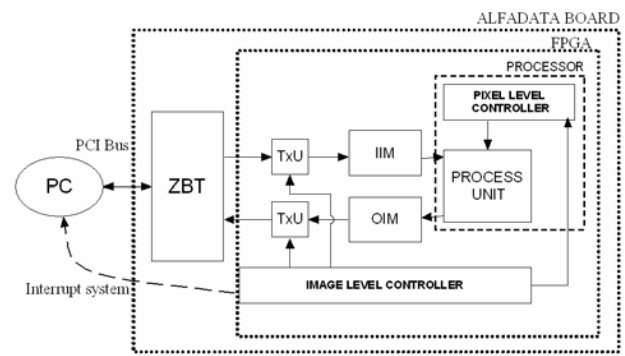

Fig. 2: Architecture general diagram

The architecture description starts with an overview of the memory management. Later, the dataflow control and the whole system operation are described and finally the processing unit is analyzed in depth.

## 3.1 Memory Management

The ZBT on-board memory size permits to store two input and one output image structure of either image type employed, QCIF (i.e. 176 x 144 pixels, approx. 200 kBytes) or CIF (i.e. 352 x 288 pixels, approx. 800 kBytes). Since the memory width is 32 bits and the pixel size is 64 bits (i.e. 8 bits per Y,U,V channels and 16 bits per Alfa and Aux channels) two memory positions are required to store one pixel. The AddressEngine coprocessor stores the upper and the lower part of the pixel in the same position of two different ZBT banks. In that way it is possible to access any pixel within only one memory cycle deploying a very simple addressing mode.

The communication between PC and the board is interrupt oriented and realized through DMA transfers. The same design style is employed both for the input and the output interfaces of the coprocessor to optimize the PC memory on-board coupling.

The whole input image is not transferred in one pass but it is divided into parts which are written to alternate ZBT blocks. Thus an optimized usage of the PCI bus is obtained and it is possible to start processing although the input image is not completely stored in the memory. The design decisions related to the size and shape of these parts are determined by software constraints, due to the AddressLib functionality. Firstly, as the pixel addressing within the input image is not random but sequential, the image is transferred in strips, horizontal or vertical depending on the way of scanning the image. The selected strip size is sixteen lines, as the maximum range of input data required to process one pixel is nine lines. The choice of a power of two number guarantees an easy addressing mode. Sixteen is also divisor of the image size, so it makes the system management easier.

In the output interface, the same idea of alternate blocks is used, to keep the optimized use of the PCI bus, but with some differences. Firstly, the upper and the lower part of each pixel are stored sequentially in the same memory bank, in such a way the PC gets the pixel data properly ordered. And secondly, the bank switching is performed only once, as soon as it is possible to start transferring the resulting image.

The expounded way of storing pixels and how the information is transferred between PC and the coprocessor results in a specific memory distribution, see figure 3. Related to this system temporal behaviour, the strip stored in the so-called *block_A* is processed while the next strip is transferred to *block_B* and vice versa. And the *Res_block_A* can be transferred when the PCI bus is free, i.e. when the input image is completely stored in the ZBT.

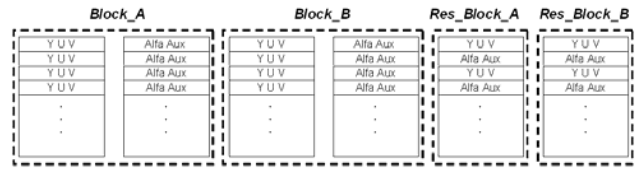

Fig. 3: ZBT memory distribution.

Relating to the intermediate memory system, the second module defined in the AddressEngine, two memory structures have been implemented. These are the so-called *IIM*, *input intermediate memory*, and the *OIM*, *output intermediate memory*.

The *IIM* is implemented at the input of the processing unit because there is a successive pixel reuse at this point of the system. Thus loading the complete neighbourhood for each pixel is avoided. Furthermore, with the implementation employed the whole neighbourhood can be obtained in only one cycle, even in the worst case with perpendicular neighbourhood and scan direction, as illustrated in figure 4. In case the neighbourhood is not loaded in the *IIM* the image level controller takes care of halting the system until it is available. The *IIM* size, in tune with the strip size, is sixteen lines. The *IIM* structure is made up of sixteen memory blocks, with two banks for the lower and the upper part of the pixel. These 32 memory blocks are implemented in the FPGA embedded memory.

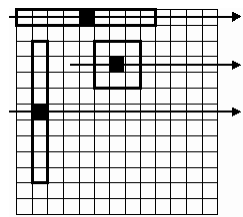

Fig. 4: Different neighbourhood types: Worst case maximum size perpendicular to scan direction

At the output of the processing unit an intermediate memory system is required as well. The *OIM* has exactly the same structure as the *IIM*, but it is needed because of different reasons. It is used as a buffer structure because there are different speeds at the interface processor unit output - ZBT memory, since the processing unit provides pixels in twice the speed than can be written to the ZBT memory.

### 3.2 Dataflow Control

According to the dataflow control, three controller blocks are implemented: The pixel level controller, the transmission unit and the image level controller.

The pixel level controller generates the control signals for the processing unit and the processing unit manages the image scanning. So the pixel level controller deals with the dataflow control between the intermediate memory system and the processing unit.

The transmission unit controls the transfer of lines from the ZBT memory to the intermediate memory system, in both the *OIM-* and the *IIM* structure.

The image level controller deals with the interrupt generation and manages as well all control blocks. So it controls the data transfers between PC and the coprocessor.

Finally, the dataflow through the system is managed by the image level controller according to the processing unit needs and requirements.

### 3.3 Interface Processor to Image Level Ctrl

The interface of the processor with the image level controller consists of a set of signals to carry out the correct supply and processing of the data.

Since the *IIM* and *OIM* act as FIFOs, the signals to control if both are in the correct state to supply and store the data to and from the processing unit consist of FULL and EMPTY signals for each of them.

For the inter addressing mode the *IIM* will take the form of two FIFOs, one for every input image, with 8 lines each. In this case we will generate the same signals for both of the FIFOs.

In order to stop the processing unit when there is no data available to read or no empty space in the OIM to write, the image level controller will disable the pixel level controller which will not proceed with any more pixel-cycles until this signal is enabled again.

### 3.4 Pixel Level Controller

The pixel level controller is the controlpath of the processor. Its purpose is to control the process unit (i.e. datapath) enabling the intervention of its components when necessary.

The datapath of our design is divided into four stages. In order to generate a result pixel one instruction has to be performed in each one of the stages. The pixel level controller (PLC from now on) will take charge of generating these instructions and executing them in the proper order.

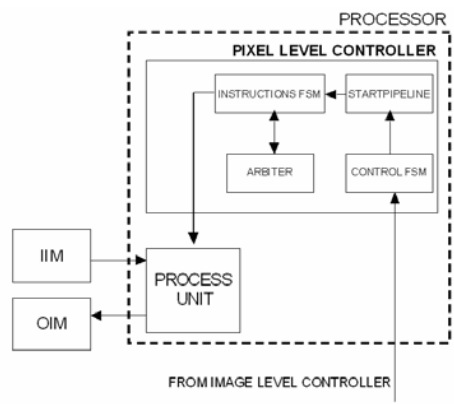

Fig. 5: Detail of the PLC inside the Processor

The PLC (fig. 5) is compound by four modules: The arbiter, the instructions FSM, the startpipeline and the control FSM (i.e. Finite-State-Machine). The control FSM generates the set of instructions to be performed in every pixel-cycle. The arbiter makes sure that the instructions in the different stages will not access to the same resources in the Process Unit. The instructions FSM can request and lock the resources in the Process Unit and generate the signals that steer the correct behavior of these resources. Finally, the startpipeline deals with the correct order of the execution of the instructions allowing us also to have instructions of different pixel-cycles in the different stages of the Process Unit being not needed to wait till one pixel-cycle is finished to start with the next one.

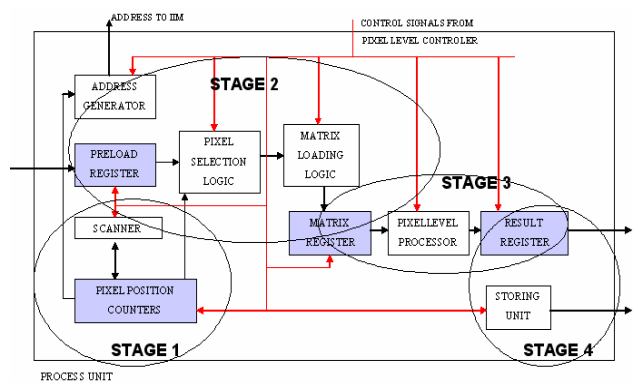

Fig. 6: Detailed structure of the Process Unit

### 3.5 Process Unit

The Process Unit (fig. 6) is the datapath of the Processor. As has been stated before, the Process Unit is divided into four different stages.

Stage 1 performs the scanning through the image. The central position of the neighbourhood in the input image for the next pixel-cycle is calculated here. Instructions that work with the pixel position counters which indicate the current pixel of the image we are sweeping belong to this stage.

All the instructions that obtain data from the *IIM* and save it to the matrix register belong to stage 2. In the matrix register is stored the whole neighbourhood that will be input for the next stage. These instructions are divided into two sets: LOAD instructions and SHIFT instructions depending on whether they fill the whole matrix from scratch or whether they only add some pixels shifting the pixels that were already in the matrix.

Stage 3 performs the execution of the pixel-operation on the neighbourhood. Operations like gradient, histogram, different filterings, etc. are carried out here.

In the fourth Stage the result pixel obtained in the third one is stored in the *OIM*.

## 4. Prototype Performance Results

### 4.1 Implementation Cost and Clock Rate

The system has been implemented with ISE 6 in a Virtex II 2V3000 board, experimental results regarding complexity and performance are shown in table 1.

Device utilization summary:

| Selected Device : 2v3000ff1152-5 | | | |
|---|---|---|---|
| Number of Slices: | 564 | out of 14336 | 3% |
| Number of Slice Flip Flops: | 216 | out of 28672 | 0% |
| Numb er of 4 input LUTs: | 349 | out of 28672 | 1% |
| Number of bonded IOBs: | 60 | out of 720 | 8% |
| Number of BRAMs: | 29 | out of 96 | 30% |
| Number of GCLKs: | 1 | out of 16 | 6% |

Timing Summary:

Minimum period: 9.784ns (Maximum Frequency: 102.208MHz)

Table. 1

Assuming that the performance of the design is constraint by the bandwidth of the PCI bus (66 MHz) which happens to be the bottleneck of the system, the clock speed should be fast enough to work with the FPGA. With this clock frequency a 264 Mbytes/s rate can be achieved between every one of the 6 ZBT RAM banks and the FPGA.

In fact, the effect in the timings due to the processing is insignificant except for some special inter operations where we cannot start processing until both of the images have been completely transferred. Even in this situation the time wasted not due to the PCI transferences is a 12.5% of the time needed to transfer the images to the board.

The high amount of block RAM used (BRAM) in the design is due to the *IIM* and *OIM* memories. In any case there is enough free memory for a possible extension of the design with other addressing schemes.

### 4.2 Performance of the Memory Architecture

In order to estimate the improvement in performance added by the memory scheme used in the hardware solution a comparison between the number of memory access operations carried out by the software solution and those made by the processor in the design has been made. Table 2 shows the results.

| Memory Accesses | | | Software solution memory accesses | Hardware solution memory accesses | Saving |
|---|---|---|---|---|---|
| Addressing | Input channels | Output channels | | | |
| Inter | Y | Y | 304.128 | 202.752 | 33% |
| Intra CON_0 | Y | Y | 202.752 | 202.752 | 0% |
| Intra CON_8 | Y | Y | 405.504 | 202.752 | 50% |
| Intra CON_8 | Y,U,V | Y,U,V | 608.256 | 202.752 | 200% |

Table 2

In table 2 CON_0 means that the one pixel neighborhood has been used and CON_8 refers to the squared 8-pixels neighborhood that appears in Figure 4.

It can be deduced from these numbers that the benefit obtained by using the memory architecture described above increases with the amount of data traffic. This is a consequence of the parallel access to the pixels in the hardware design. There, all the channels of the new pixels needed to update the neighborhood are loaded in parallel while in the software solution this is done sequentially. Therefore a structure like this is beneficial for the addressing of pixel neighborhoods.

### 4.3 Comparison With a Software Implementation

The performance of the hardware solution has been tested against a software platform. This test consisted of an execution of the MPEG-7 Global Motion Estimation Software [6] over 4 different test sequences on a Pentium Mobile (PM) at 1.6 GHz with 512 MB DDR RAM.

The hardware platform consisted of the ADM-XRCII board connected to the PCI bus of a PC with a Pentium 4 microprocessor running at 3 GHz. The top-level software layer of the Global Motion Estimation Software was kept in the PC, which accessed to the ADM-XRCII board after every call to the AddressLib.

This global motion estimation software is used for Mosaicing purposes, the test sequences are MPEG-1 CIF-sized and as a result this software creates a Mosaic with the global motion of the scene.

Table 3 shows that the prototype implementation running with 66 MHz improves performance by an average factor of 5 over a PM running with 1.6 GHz. The bottleneck is the bandwidth of the PCI bus, which restricts the loading of the images. In future implementations, the PC in fig. 2 could be substituted by an embedded RISC, e.g. PowerPC on the FPGA and the PCI bus replaced by a more powerful on-chip bus, e.g. CoreConnect.

| Video | Time in PM | Time in FPGA | Intra AddrEng calls | Inter AddrEng calls |
|---|---|---|---|---|
| Singapore | 4'35'' | 1'04'' | 4542 | 3173 |
| Dome | 5'28'' | 1'13'' | 4931 | 3404 |
| Pisa | 12'25'' | 2'21'' | 9294 | 6541 |
| Movie | 5'22'' | 1'05'' | 4070 | 3085 |

Table 3

## 5. Outlook

The implementation of intra and inter addressing schemes in an FPGA was described in this paper. The next step will be to implement the segment addressing scheme on the same FPGA board. Further on there are two directions: 1) Implementation in standard cell ASIC for further power and performance optimization. 2) Exploitation of dynamically reconfigurable FPGAs for more complex processing in research.

For exploitation of dynamic reconfigurability, an FPGA with embedded RISC core and partial dynamic reconfiguration capabilities will be used. The pixel addressing will be implemented in a statically configured block of the FPGA, as all supported algorithms are using the same AddressLib scheme, whereas the pixel processing, which might be changed during the process of video analysis, will be implemented in a dynamically reconfigurable block. High level control operations will be implemented on the embedded RISC processor.

## 6. Summary

The focus of this paper is on the design of a coprocessor architecture for visual information processing, where pixel addressing has shown to be the dominant operation, exceeding even pixel processing. Therefore pixel addressing was in the focus of optimization. Structured pixel addressing schemes have been implemented both in a software library, called AddressLib, and in a coprocessor architecture, called AddressEngine, described in this paper.

The first version of the AddressEngine was implemented on a FPGA board. A set of controllers was defined to control the dataflow within the AddressEngine and to/from the PC. A pipelined architecture for pixel processing and address calculations was developed. Memory accesses are in parallel on complete neighbourhoods of pixels. This first version of the AddressEngine supports intra and inter addressing schemes. Segment addressing is planned for future versions.

The performance of the AddressEngine was analyzed regarding processing speed and circuit complexity on the FPGA. A comparison with a software implementation of the MPEG-7 Global Motion Estimation algorithm has shown that our prototype achieves an average speedup factor of 5, if the high level algorithm is kept fully programmable on the main CPU.

## Acknowledgment

This material is based upon work supported by the IST program of the EU in the project IST-2000-32795 SCHEMA (http://www.iti.gr/schema)

## References

[1] V.Mezaris et al: "A test-bed for region-based image retrieval using multiple segmentation algorithms and the MPEG-7 eXperimentation Model: The Schema Reference System." CIVR 2004, International Conference on Image and Video Retrieval, Dublin, July 21-23, 2004

[2] S. Herrmann et al: "A Video Segmentation Algorithm for Hierarchical Object Representations and Its Implementation", IEEE Transactions on Circuits and Systems for Video Technology, Vol. 9, No. 8, Dec. 1999, pp. 1204-1215

[3] W. Stechele et al: „Towards a Dynamically Reconfigurable System-on-Chip Platform for Video Signal Processing", International Conference on Architecture of Computing Systems, ARCS 2004, Augsburg, Germany, March 23-26, 2004

[4] J. Becker: „Configurable Systems-on-Chip: Commercial and Academic Approaches", ICECS 2002, Dubrovnik, September 2002

[5] K. Compton, S. Hauck: „Reconfigurable Computing: A Survey of Systems and Software", ACM Computing Surveys, Vol. 34, No. 2, pp. 171-210, June 2002

[6] MPEG-7 eXperimentation Model (XM) http://www.lis.ei.tum.de/research/bv/topics/mmdb/mpeg7.html